\documentclass[11pt]{article}
\usepackage{color}
\usepackage{amssymb}
\usepackage{amsmath}
\usepackage{algorithm}
\usepackage[linesnumbered, figure, algo2e]{algorithm2e}
\usepackage{url}
\usepackage{graphicx}

\usepackage{subfigure}
\usepackage{tikz}

\usepackage{color}

\usepackage{paralist}
\usepackage[margin=1in]{geometry}

\newcommand{\proof}{\textit{Proof: }}
\newcommand{\complexity}[1]{\text{$O(#1)$}}
\newcommand{\var}[1]{\text{$#1$}}
\newcommand{\func}[1]{{\textnormal{\texttt{#1}}}}
\newcommand{\comment}[1]{\textnormal{\##1}}
\newcommand{\keyword}[1]{{\textnormal{\textbf{#1}}}}

\newtheorem{theorem}{Theorem}
\newtheorem{problem}{Problem}

\title{CacheDiff: Fast Random Sampling}
 \author{Dai Bui}

\begin{document}

\maketitle
\begin{abstract}
We present a sampling method called, CacheDiff, that has both time and space complexity of \complexity{k} to randomly select \var{k} items from a pool of \var{N} items, in which \var{N} is known. 
\end{abstract}
\section{Introduction}
In this paper, we study the following problem:
\begin{problem}\label{prob:selectKfromN}
Select \var{k} items from a pool of given \var{N} items uniformly.
\end{problem}

This problem has been studied extensively in~\cite{Vitter1985ReservoirSampling, Efraimidis2006WRS, Chao1982UPS, Meng2013ScalableRandomSampling, Vitter1984FasterMethodsRandomSampling,Ahrens1985SequentialRandomSampling}. The applications of this problem span from security to big data. However, the approaches in~\cite{Vitter1985ReservoirSampling, Efraimidis2006WRS, Chao1982UPS} have time complexity of \complexity{N}, which is not very efficient to use to sample in big data when \var{N} is often very big or to generate random codes whose probability to be guessed is extremely small. In~\cite{Vitter1984FasterMethodsRandomSampling}, Vitter presented an acceptance-rejection method to sequentially select \var{k} items from a pool of given \var{N} items uniformly with complexity of approximately \complexity{k} when \var{k} is very small compared to \var{N}. The experiments in~\cite{Vitter1984FasterMethodsRandomSampling} shows that when $\var{k} \geq 0.15\var{N}$, the running time of the acceptance-rejection method is worse than the reservoir sampling in~\cite{Vitter1985ReservoirSampling}. In particular, the acceptance-rejection method only works with known \var{N}.

Random sampling algorithms are useful in several areas:
\begin{itemize}
\item Big data: Instead of processing all data items, we can process only process a subset of the items to obtain approximate results. The subset of the items can be selected by randomly sample \var{k} items from all the data items. 
\item Election polling: To estimate the approximation approval rate of election candidates, instead of conducting a full survey, we can randomly select people to obtain their opinions on candidates.
\item Online tickets: Each user when buying an online ticket will be generated a code that is very difficult to predict by hackers. To avoid the code duplication for different tickets, the codes are generated in batches by selecting randomly \var{k} integers (each selected integer is a code) from the integers between 0 and \var{N}. To make the codes very difficult to predict, then \var{k} need to be very small compared to \var{N}.
\end{itemize}

It is rather straightforward to see that Problem~\ref{prob:selectKfromN} can be solved using a random permutation (shuffling) algorithm.

\begin{minipage}[t]{0.49\textwidth}
\null 
 \begin{algorithm}[H]
\SetAlgoVlined
\SetKwFunction{algo}{randomSampling}
\SetKwProg{myalg}{function}{}{}
\myalg{\algo{\var{a},\var{N},\var{k}}}{
\For{$\var{i}=\var{N}-1\ \keyword{to}\ \var{N}-\var{k}$}{
	$\var{j}=\func{random}(\var{0},\var{i})$
	
	\comment{exchange \var{a}[\var{i}] and \var{a}[\var{j}]}
	
	$\var{t}= \var{a}[\var{i}]$
	
	$ \var{a}[\var{i}] =  \var{a}[\var{j}]$
	
	 $\var{a}[\var{j}] = t$
	
}
\Return $\var{a}[(\var{N}-\var{k})..(\var{N}-1)]$
}
\caption{A Simple Random Sampling}
\label{alg:randomShuffling}
\end{algorithm}
\vspace*{-0.5cm}
\begin{algorithm}[H]
\SetAlgoVlined
\SetKwFunction{algo}{randomIndexSampling}
\SetKwProg{myalg}{function}{}{}
\myalg{\algo{\var{N},\var{k}}}{
$\var{index}=\func{vector}(N)$\comment{allocate index array}

\For{$\var{i}=0\ \keyword{to}\ \var{N}-1$}{
	$ \var{index}[\var{i}] = i$ \comment{initialize index array}
}

\For{$\var{i}=\var{N}-1\ \keyword{to}\ \var{N}-\var{k}$}{
	$\var{j}=\func{random}(0,\var{i}) $
	
	\comment{exchange \var{index}[\var{i}] and \var{index}[\var{j}]}
	
	$\var{t}= \var{index}[\var{i}]$
	
	$ \var{index}[\var{i}] =  \var{index}[\var{j}]$
	
	 $\var{index}[\var{j}] = t$
	
}
\Return $\var{index}[(\var{N}-\var{k})..(\var{N}-1)]$
}
\caption{Initial Random Index Sampling}
\label{alg:randomIndexSelection}
\end{algorithm}
\end{minipage}%
\hspace{0.02\textwidth}
\begin{minipage}[t]{0.49\textwidth}
\null
\begin{algorithm}[H]
\SetAlgoVlined
\SetKwFunction{algo}{cacheDiffRandomIndexSampling}
\SetKwProg{myalg}{function}{}{}
\myalg{\algo{\var{N},\var{k}}}{

$\var{me}=\func{hash\_table}()$

$\var{output}=\func{vector}(\var{k})$

\nl\For{$\var{i}=\var{N}-1\ \keyword{to}\ \var{N}-\var{k}$}{\label{ln:cacheDiffForLoop}
\nl	$\var{j}=\func{random}(0,\var{i})${\label{ln:cacheDiffRandomGen}}
	
	\comment{exchange \var{index}[\var{i}] and \var{index}[\var{j}]}
	
	\eIf{\var{me}.\func{has\_key}\textnormal{(\var{j})}} {
		\var{index\_j}=\var{me}[\var{j}]
	}{
		\var{index\_j}=\var{j}
	}
	
\nl	\eIf{\var{me}.\func{has\_key}\textnormal{(\var{i})}} {
\nl		\var{index\_i}=\var{me}[\var{i}]
	}{
\nl		\var{index\_i}=\var{i}
	}
	
\nl	\var{me}[\var{i}]=\var{index\_j}
	
	\var{me}[\var{j}]=\var{index\_i}
	
	\var{output}.\func{push\_back}(\var{index\_j})
}
\Return \var{output}
}
\caption{CacheDiff Random Index Selection}
\label{alg:cacheDiffRandomIndexSelection}
\end{algorithm}
\end{minipage}

Algorithm~\ref{alg:randomShuffling} above requires time and space complexity of \complexity{N}, which can be prohibitive in big data or highly secure random code generators. However, note that Problem~\ref{prob:selectKfromN} is equivalent to the following problem:

\begin{problem}\label{prob:selectIndexKfromN}
Select \var{k} unique integers, e.g., indices of the items, from the integers between 0 and \var{N}-1 uniformly.
\end{problem}


\begin{figure}[h]
\includegraphics[width=\textwidth]{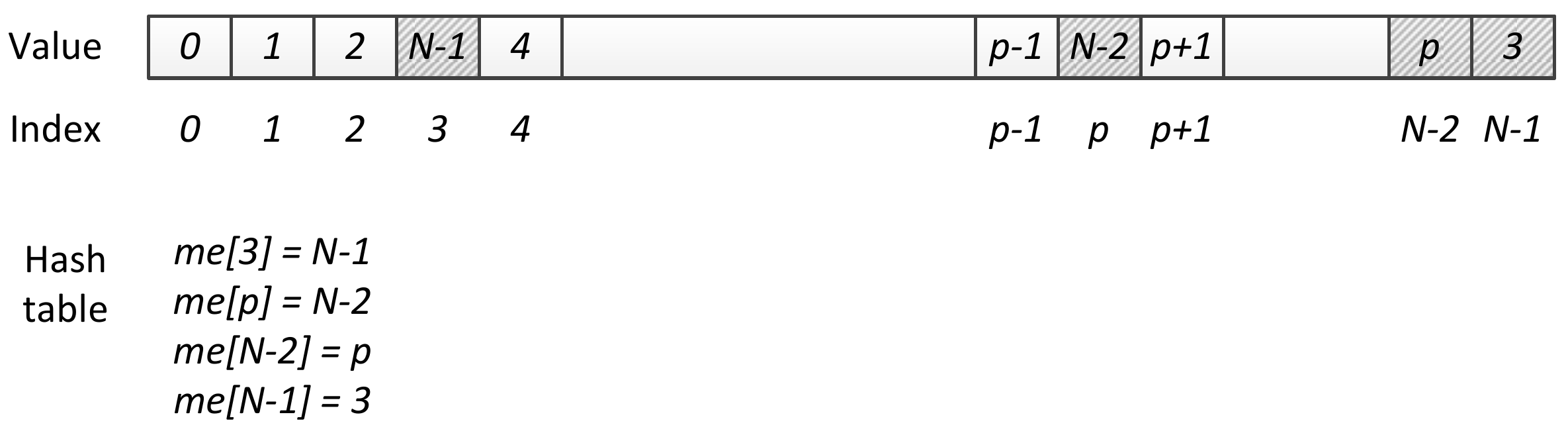}
\caption{CacheDiff Technique to Selectively Store a Small Number of Modified Entries in a Large Array}
\label{fig:simpleCacheDiff}
\end{figure}

Algorithm~\ref{alg:randomIndexSelection} solves Problem~\ref{prob:selectIndexKfromN}. Not that Algorithm~\ref{alg:randomIndexSelection} is similar to Algorithm~\ref{alg:randomShuffling}. 
Algorithm~\ref{alg:randomIndexSelection} still requires \complexity{N} time and space. However, note that when \var{k} is very small compared to \var{N}, $\var{k}\ll\var{N}$, $\var{index}[\var{i}]=\var{i}$ for most of \var{i}. Because the array of integers from 0 to (\var{N}-1) can be stored very efficiently, as a result, to store the \var{index} array, we only need to \textit{cache} the value in the \var{index} array that $\var{index}[\var{i}]\neq\var{i}$. We can implement this caching using a hash table. As a result, we can improve the time and space complexity of Algorithm~\ref{alg:randomIndexSelection} using Algorithm~\ref{alg:cacheDiffRandomIndexSelection}.

\section{CacheDiff Random Sampling}\label{sec:cacheDiffRandomSampling}

Algorithm~\ref{alg:cacheDiffRandomIndexSelection} improves from Algorithm~\ref{alg:randomIndexSelection} in both time and space complexity by using a hash table to store the difference between the output array and the simple array of integers from 0 to \var{N}-1. Because we only select \var{k} items, as a result, the space complexity of the hash table \var{me} is \complexity{\var{k}}. Then it is easy to see that Algorithm~\ref{alg:cacheDiffRandomIndexSelection} runs in \textit{average} time complexity of \complexity{\var{k}} and requires \complexity{\var{k}} space.

Figure~\ref{fig:simpleCacheDiff} illustrates the CacheDiff technique. In the first iteration, 3 is selected so we swap 3 and \var{N}-1. The entries at 3 and \var{N}-1 now become different from the indices so the hash table caches the values at those entries. Similarly for the second iteration when \var{p} is selected. 

\begin{theorem}
Each index from 0 to \var{N-1} has the same probability of $\frac{\var{k}}{\var{N}}$ to be selected by Algorithm~\ref{alg:cacheDiffRandomIndexSelection}.
\end{theorem} 

\proof First, we prove that the probability that an index is \textit{not} selected after \var{n} iterations of the for loop at line~\ref{ln:cacheDiffForLoop} is $\frac{\var{N}-\var{n}}{\var{N}}$. We will prove it using induction. For $\var{n}=1$, then $\var{i}=\var{N}-1$, as a result, the probability that the index is selected is $\frac{1}{\var{N}}$. Suppose that our hypothesis holds for $\var{n}=\var{m}$, we will prove that it holds for $\var{n}=\var{m}+1$. At iteration $\var{n}=\var{m}+1$, $\var{i}=\var{N}-\var{m}-1$. Then the probability that the index is selected due to the random selection at line~\ref{ln:cacheDiffRandomGen} is $\frac{1}{\var{N}-\var{m}}$. Then the probability that the index is \textit{not} selected \textit{at} iteration $\var{n}=\var{m}+1$ is $1-\frac{1}{\var{N}-\var{m}}=\frac{\var{N}-(\var{m}+1)}{\var{N}-\var{m}}$. As a result, the probability that the index is \textit{not} selected after $\var{n}=\var{m}+1$ iterations is $\frac{\var{N}-\var{m}}{\var{N}}\times\frac{\var{N}-(\var{m}+1)}{\var{N}-\var{m}}=\frac{\var{N}-(\var{m}+1)}{\var{N}}$, which is what we want to prove.

So after \var{k} iterations, the probability that one index is not selected is $\frac{\var{N}-\var{k}}{\var{N}}$, then the probability that the index is selected in one of the \var{k} iteration is $1-\frac{\var{N}-\var{k}}{\var{N}}=\frac{\var{k}}{\var{N}}$.$\hfill\Box$ 

\section{Conclusions}
In this paper, we demonstrated the use of a hash table to store a small number of modified entries within a predictable sequence, in the words of information theory~\cite{Cover2006ElementInfoTheory}, the sequence has a small entropy. This method lead to a simple algorithm that has lower complexity than~\cite{Vitter1985ReservoirSampling, Efraimidis2006WRS, Chao1982UPS, Meng2013ScalableRandomSampling,Ahrens1985SequentialRandomSampling} or easier to understand and implement than~\cite{Vitter1984FasterMethodsRandomSampling}.

\bibliographystyle{plain}
\bibliography{CacheDiffSampling}
\end{document}